%% file: main.tex
\documentclass[aps,prl,reprint,superscriptaddress]{revtex4-2}
\usepackage{graphicx}
\usepackage{amsmath}
\usepackage{amssymb}
\usepackage{mathrsfs}
\usepackage[T1]{fontenc}
\usepackage{calligra}
\usepackage{array}
\usepackage{xcolor}
\usepackage{float}
\usepackage{soul}
\usepackage{bm}
\usepackage[toc,page]{appendix}
\usepackage{braket}
\usepackage{wrapfig}
\usepackage[colorlinks=true, linkcolor=blue, urlcolor=blue,citecolor=blue]{hyperref}

\hypersetup{colorlinks,linkcolor=blue,citecolor=blue,urlcolor=blue}

\begin{document}
    %------------------
    %% Title and authors
    %------------------
	\title{Spin Hall Effect: Symmetry Breaking, Twisting, and Giant Disorder Renormalization}
	
	\author{David T. S. Perkins}
    \email[]{david.t.s.perkins@york.ac.uk}  
	\affiliation{School of Physics, Engineering and Technology and York Centre for Quantum Technologies, University of York, YO10 5DD, York, United Kingdom}

	\author{Alessandro Veneri}
	\affiliation{School of Physics, Engineering and Technology and York Centre for Quantum Technologies, University of York, YO10 5DD, York, United Kingdom}
	
	\author{Aires Ferreira}
	\email[]{aires.ferreira@york.ac.uk}
	\affiliation{School of Physics, Engineering and Technology and York Centre for Quantum Technologies, University of York, YO10 5DD, York, United Kingdom}
	
    %------------------
    %% Abstract
    %------------------
	\begin{abstract}
    Atomically-thin materials based on transition metal dichalcogenides and graphene offer a promising avenue for unlocking the mechanisms underlying the spin Hall effect (SHE) in heterointerfaces. Here, we develop a microscopic theory of the SHE for twisted van der Waals heterostructures that fully incorporates twisting and disorder effects, and illustrate the critical role of symmetry breaking in the generation of spin-Hall currents. We find that an accurate treatment of vertex corrections leads to a qualitatively and quantitatively different SHE than that obtained from popular approaches like the ``$i\,\eta$'' and  ladder approximations. A pronounced oscillatory behavior of skew-scattering processes with twist angle, $\theta$, is predicted, reflecting a non-trivial interplay of Rashba and valley-Zeeman effects and yields a vanishing SHE for $\theta = 30^\circ$ and, for    graphene-WSe$_2$, an optimal SHE for $\theta \approx 17^\circ$. Our findings reveal disorder and broken symmetries as important knobs to optimize interfacial SHEs.
	\end{abstract}
	\maketitle
            
    %------------------
    %% Intro
    %------------------
    
    % Twist focused
    The discovery of superconductivity \cite{Cao2018,Isobe2018}, flat bands \cite{Shallcross2010,Bistritzer2011,Trambly2012,Moon2018}, strongly-correlated insulating phases and topological behavior \cite{Geim2013,Novoselov2016} in layer-twisted honeycomb systems has lead them to be at the centre of many theoretical and experimental studies \cite{Zou2018,Choi2019,Classen2019,Lu2019,Lian2019,Tarnopolsky2019,Yankowitz2019,Cao2020,Padhi2020,Liao2021,Song2022,Duan2022}. The significance of twisting in the plethora of spin-dependent phenomena generated by spin-orbit coupling (SOC) is currently under intense investigation \cite{Li2019,David2019,Peterfalvi2022,Pezo_2021,Naimer_2021,Veneri2022,Lee2022,Sun_23,Rao_23}. It has been shown that the modification of the Fermi surface's spin texture induced by twisting \cite{Li2019,David2019,Peterfalvi2022,Pezo_2021,Naimer_2021} leads to profound changes in the spin-charge interconversion processes displayed by graphene-transition metal dichacolgenide (TMD) bilayers \cite{Veneri2022,Lee2022}, a paradigmatic system in the burgeoning field of graphene spintronics \cite{Avsar2020,Sierra2021,Perkins2024}. 
    Despite these advances, a microscopic theory of the spin Hall effect (SHE) -- the generation of a transverse spin current due to an applied electric field -- reflective of the untwisted, let alone twisted, van der Waals (vdW) heterostructures used in  spin Hall experiments \cite{Avsar2014,Wang2015,Ghiasi2017,Benitez2018,Safeer2019,Ghiasi2019,Benitez2020,Li2020,Hoque2021} remains an elusive task. Such a theory could offer valuable insight into the role of broken spatial symmetry and relative atomic orientation between layers. Through twisting, the single unique mirror plane present in aligned graphene-TMD bilayers is lost, reducing the symmetry from $C_{3v}^{\null}$ to the chiral point group $C_{3}^{\null}$. From a physical perspective, the metal-chalcogen environment around each carbon atom is changed as the layers are twisted, leading to a modulation of the out-of-plane asymmetry SOC (Rashba) and the sublattice-resolved SOC (valley-Zeeman). Twisted vdW heterostructures therefore provide a natural, highly tunable platform to investigate interfacial SHEs and may serve as a guide in examining other heterointerfaces.
    
    Another key question is the breaking of translation symmetry due to disorder, which is known to profoundly modify the electrodynamic response of spin-orbit-coupled Dirac bands \cite{Perkins2024}. The  ubiquitous nature of   disorder in 2D crystals makes it a crucial ingredient for understanding both the SHE and the wealth of magneto-electric effects underlying charge-to-spin conversion, such as the inverse spin galvanic effect (ISGE). The ISGE has been understood in both untwisted \cite{Offidani2017,Sousa2020} and twisted \cite{Veneri2022} 2D vdW heterostructures with dilute random impurities. In contrast, previous theoretical work on the SHE has focused on minimal models of proximitized  graphene, i.e. without disorder \cite{Drydal2009,Zollner_23},  within the Rashba spin gap \cite{Monaco2021}, and in the absence of the valley-Zeeman effect \cite{Ferreira2021}.  The diffusive SHE with a Fermi energy located well above the spin gap -- the most experimentally accessible and well controlled regime due to the suppression of carrier-density inhomogeneities   \cite{Avsar2014,Wang2015,Ghiasi2017,Benitez2018,Safeer2019,Ghiasi2019,Benitez2020,Li2020,Hoque2021} -- is theoretically  challenging, and more so for comprehensive graphene-TMD models with competing symmetry-breaking effects.  Unlike the ISGE, where the nonequilibrium spin density is simply proportional to the charge transport time, the extrinsic SHE is governed by its own time scales (which, technically speaking, are encoded in  vertex corrections to spin-charge response functions). The microscopic processes governing the SHE reflect the rich interplay between Fermi-surface spin texture (quantum geometry) and spin-orbit scattering mechanisms (extrinsic effects), and hence constitute a critical puzzle piece in understanding  non-local spin transport experiments \cite{Benitez2018,Safeer2019}, as well as guiding future efforts in spin-twistronics. While real-space techniques for billion-atom calculations of linear response functions have recently been devised  that could in principle be used to overcome the aforementioned challenges \cite{Ferreira_15,Kite_20,Castro_24}, the scarce numerical studies of SHE  in the literature simulated small  systems (of order 10 nm in linear size,  i.e. quantum dot territory \cite{Garcia_17}), and hence are far from capturing  the transport regimes seen in experiments.
    
    In this Letter, we construct a microscopic theory for twisted graphene-TMD systems that accounts for band structure effects non-perturbatively and straddles strong and weak scattering regimes, hence overcoming the above challenges via a unified approach. The most surprising result is a giant modulation of the spin Hall conductivity with twist angle, yielding an optimal SHE for chiral bilayers at a critical twist angle ($\theta_c \approx 17^\circ$ for graphene-WSe$_2$). %and a small  SHE around $\theta = 30^\circ$ due to the symmetry-enforced vanishing of valley-Zeeman effect. 
    This novel behavior, reflective of the sensitivity of disorder  corrections to quantum-geometric effects, is absent in the ``$i\,\eta$'' approximation.
    %, which fails   qualitatively in predicting the    steady-state SHE.
    %Here, we expect spatial fluctuations of the proximity-induced SOCs to dominate the SHE. Finally, to obtain spin Hall angles in-line with observation, we discern the need for quantum corrections beyond semiclassical transport. 
    Moreover, our findings   suggest that purely diffusive SHEs in graphene-TMD  are dominated by skew-scattering processes with large cross sections. An intriguing exception are $C_{3v}$-invariant systems with $\theta =  30^\circ$. Here, anomalous scattering processes  due to spatial fluctuations of the proximity-induced SOCs \cite{Drydal2012,Huang_16,Milletari2016} are expected to govern the steady-state SHE.
    
    %------------------
    %% Model
    %------------------
    
    \textit{Model and Theory.---}We implement the  Hamiltonian of Refs. \cite{Li2019,David2019,Peterfalvi2022,Pezo_2021,Naimer_2021,Veneri2022} for the low-energy graphene-TMD  description, which assumes the axes to be taken in the graphene sheet's frame of reference. Specifically ($\hbar = 1$), for the clean system we write $H_{\mathbf{k}} = H_{0\mathbf{k}} + H_{\text{R}} + H_{\text{vz}}$ ($\textbf{k}$ is the wavevector measured from a Dirac point), with $H_{0\mathbf{k}} = v (\tau_{z} \sigma_{x} k_{x} + \sigma_{y} k_{y})$, $H_{\text{vz}} = \lambda_{\text{vz}}(\theta) \tau_{z} s_{z}$, and
    \begin{equation}
    	H_{\text{R}} = \lambda_{\text{R}}(\theta) e^{i s_{z} \frac{\alpha_{\text{R}}(\theta)}{2}} (\tau_{z} \sigma_{x} s_{y} - \sigma_{y} s_{x} ) e^{-i s_{z} \frac{\alpha_{\text{R}}(\theta)}{2}},
    	\label{Rashba_Hamiltonian}
    \end{equation}
    where $\lambda_{\text{R}}(\theta)$, $\alpha_{\text{R}}(\theta)$, and $\lambda_{\text{vz}}(\theta)$ are the twist-dependent \textit{Rashba magnitude}, \textit{Rashba phase}, and \textit{valley-Zeeman coupling}, respectively, $v$ is the bare Fermi velocity, and $\tau_{i}$, $\sigma_{i}$, and $s_{i}$ ($i \in \{x,y,z\}$) are the Pauli matrices acting on the valley, sublattice, and spin degrees of freedom, respectively.  Here, we use the $\theta$ dependence of the SOC magnitudes accurately mapped by recent quantum interference measurements on twisted graphene-WSe${}_{2}$ \cite{Sun_23} to predict the full $\theta$ dependence of the SHE. Furthermore, we include scalar disorder into our model via the term $V(\textbf{r}) = \sum_{i} u_{0} \,\delta(\mathbf{r}-\mathbf{R}_{i})$, where $\{\mathbf{R}_{i}\}$ is the set of impurity positions and $u_{0}$ characterises the impurity scattering strength. Short-range disorder typically dominates the electronic transport when probing behaviour away from the charge neutrality point \cite{Peres_RMP_10,Ni_10,Joucken_21} (thus electron-hole puddles are expected to play no role in the transport physics located above the Rashba gap). We further note that the twisted bilayer system under study generally  belongs to the $C_3$ chiral group, except for the discrete set of twist angles $\theta= p \pi/3$ ($p \in \mathbb{Z}$), at which the symmetry is elevated to $C_{3v}$ due to the presence of a mirror plane. Moreover, as shown below, there is an important hidden symmetry for $\theta = \pi/6 $. These  considerations will become crucial when assessing the disorder corrections to SHE.
    
    \begin{figure}[t]
        \centering
        \includegraphics[width=\linewidth]{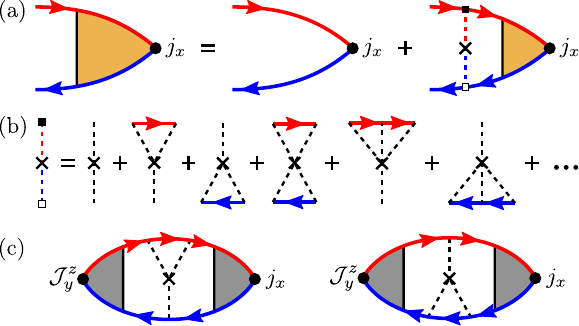}
        \caption{(a): Renormalized charge current vertex within the $T$-matrix formalism. Solid red (blue) lines denote disorder-averaged retarded (advanced) Green's functions, red (blue) dashed lines with black (white) boxes represent retarded (advanced) $T$-matrices, and the black cross signifies the insertion of the scalar impurity density. (b): Expansion of the $T$-matrix vertex renormalization. The black dashed lines denotes an impurity scattering event. (c): $Y$-diagrams. Response functions with the insertion of a single third-order scattering event. The grey shading indicates the renormalization of the vertices within the BA (retaining only the first term in Fig. \ref{Diagrammatics}b).}
        \label{Diagrammatics}
    \end{figure}
    
    The spin Hall conductivity, $\sigma^{\text{sH}}$, is calculated from the Kubo-Streda formula \cite{Streda1982,Crepieux2001,Milletari2016} using an extension of the  $T$-matrix diagrammatic technique of Ref. \cite{Milletari2016} to spin-orbit-coupled bands. The  transport in the dilute impurity regime is governed by the Fermi-surface contribution
    \begin{equation}
        \sigma^{\text{sH}} = \sum_{\mathbf{k}} \text{tr}\left[ \mathcal{J}_{y}^{z} \mathcal{G}^{+}_{\mathbf{k}} \widetilde{j}_{x} \mathcal{G}^{-}_{\mathbf{k}} \right],
        \label{SHE_TM}
    \end{equation}
    where $\mathcal{G}_{\mathbf{k}}^{\pm} = (G_{0,\mathbf{k}}^{\pm-1} - \Sigma^{\pm})^{-1}$ are the  disorder-averaged retarded ($+$)/advanced ($-$) Green's functions at the Fermi energy, $\varepsilon$, $\Sigma^{\pm}$ are the disorder self-energies, $G_{0,\mathbf{k}}^{\pm} = (\varepsilon - H_{\mathbf{k}} \pm i0^{+})^{-1}$ are the clean Green's functions, $\mathcal{J}_{y}^{z} = v \sigma_{y} s_{z}/2$ is the spin current operator, $j_{x} = -e\partial_{k_{x}} H_{\mathbf{k}}$ is the charge current operator in the $x$ direction ($e > 0$), $\widetilde{j}_{x}$ is the disorder-renormalized charge current operator (Fig. \ref{Diagrammatics}), and the trace is taken over all internal degrees of freedom. Eq. (\ref{SHE_TM}) captures all possible single-impurity scattering processes when handled within the $T$-matrix formalism outlined in Figs. \ref{Diagrammatics}(a)-(b). Most notably it accounts for skew-scattering (semiclassical) and side-jump (quantum) corrections in a fully non-perturbative fashion \cite{Milletari2016}. If vertex corrections are ignored, $\sigma^{\text{sH}}$ fails to vanish when $\lambda_{\text{vz}} = 0$, thus violating the exact SU(2)-gauge covariance of the Rashba-coupled system \cite{Dimitrova2005,Milletari2017}. As it turns out, vertex corrections are also essential when sublattice symmetry is broken, i.e. $\lambda_{\text{vz}} = \lambda_{A} - \lambda_{B} \neq0$ \cite{Perkins2024}, where $\lambda_{A(B)} $ is the intrinsic-like SOC on $A(B)$ sites. We demonstrate this in two complementary ways:  by means of a numerical  evaluation of the $T$-matrix series (full resummation) and an analytical calculation of a sub-set of Feynman diagrams. The latter provides insights into the microscopic mechanisms governing the SHE, while the former allows us to reach the strong  and unitary scattering regimes (e.g, describing resonant impurities \cite{Ferreira2011}).
    
    %-------------------------
    %% Results
    %-------------------------
    
    \textit{Results.---}We specialize to the case $|\varepsilon| > \Delta_s$, where $\Delta_s = \sqrt{4\lambda_{\text{R}}^{2}+\lambda_{\text{vz}}^{2}}$ is the spin gap, which, as mentioned previously, is the most pertinent parameter region. We start by describing the impact of skew scattering to leading order in $u_0$. This is achieved by calculating the $Y$-diagrams shown in Fig. \ref{Diagrammatics}(c), in which the Green's functions and vertices are renormalized within the first Born approximation (BA),
    \begin{equation}
       \sigma_{Y}^{\text{sH}}=\sum_{\mathbf{k},\mathbf{p}}2\,\text{Re}\left\{ \text{tr}\left[\mathcal{G}_{\mathbf{k}}^{-}\bar{\mathcal{J}}_{y}^{z}\mathcal{G}_{\mathbf{k}}^{+}Y^{+}\mathcal{G}_{\mathbf{p}}^{+}\bar{j}_{x}\mathcal{G}_{\mathbf{p}}^{-}\right]\right\} ,
        \label{SHE_Y}
    \end{equation}
    where $Y^{+} = nu_{0}^{3} \sum_{\mathbf{q}} \mathcal{G}^{+}_{\mathbf{q}}$ is the retarded skew-scattering insertion, $\mathcal{G}_{\mathbf{k}}^{\pm}$ are the Green's functions evaluated within the BA, and $\bar{\mathcal{J}}_{y}^{z}$ and $\bar{j}_{x}$ are the disorder-renormalized spin current and charge current vertices, respectively, calculated within the BA. 
    %These disorder-renormalized vertices are treated by performing a Clifford algebra decomposition \cite{Ferreira2021}. 
    We note that the Rashba phases in Eq. (\ref{Rashba_Hamiltonian}) can be removed by \textit{untwisting} the full Hamiltonian via a unitary spin rotation (see  Ref. \cite{Veneri2022} for details).

    Evaluating the $Y$-diagrams in Fig. \ref{Diagrammatics} produces
    \begin{equation}
        \sigma_{Y}^{\text{sH}} = \frac{2e\varepsilon}{n\pi u_{0}} \frac{\lambda_{\text{R}}^{4}\lambda_{\text{vz}}^{2}(\varepsilon^{2}-\lambda_{\text{vz}}^{2})(\varepsilon^{2}+\lambda_{\text{vz}}^{2})^{2}}{(\varepsilon^{4}(\lambda_{\text{R}}^{2}+\lambda_{\text{vz}}^{2}) + 3\lambda_{\text{R}}^{2}\lambda_{\text{vz}}^{4} - \varepsilon^{2}\lambda_{\text{vz}}^{4})^{2}},
        \label{SHE_Y_result}
    \end{equation}
     to leading order ($\mathcal{O}(n^{-1})$) in the impurity concentration. The intricate behavior with the  Fermi energy  and $\theta$-dependent SOCs encoded in Eq. (\ref{SHE_Y_result}) reflects a remarkable reliance of disorder effects on the spin-orbital texture of Bloch wavefunctions  (the quantum geometry of energy bands \cite{Offidani2017}). Principally, a non-coplanar Rashba spin texture (and thus $\lambda_{\text{vz}}\neq 0$) is required for a non-vanishing SHE. This has a simple interpretation: skew scattering from scalar impurities relies upon electronic states with a well-defined  spin polarization around a valley to enable a clear separation between spin-up and spin-down scattering channels. The tilted Rashba spin textures in graphene-TMD  generally satisfy this requirement. Thus, a spin Hall response naturally emerges when the sublattice symmetry is broken   (note that $\lambda_{\text{R}}(\theta)$ is guaranteed to be non-zero due to the interfacial breaking of the horizontal mirror plane). 
     These considerations remains true at $\mathcal{O}(n^{0})$, further emphasising the critical role played by vertex corrections. In addition to occurring at $\mathcal{O}(n^{-1})$, the renormalized response also carries a factor of $u_{0}^{-1}$, which puts it at the next order in the scattering strength when compared to the electrical conductivity and spin susceptibility \cite{Ferreira2021,Veneri2022}. Anomalous scattering processes, such as single-impurity side jumps and diffractive skew scattering \cite{Ado_2015,Milletari2016,Milletari_2016_2,Sinitsyn_06}, kick in to next order in the small-$n$ expansion, and thus are relevant for samples with low carrier mobility. They are not considered here.

\begin{figure}[t]
        \centering
        \includegraphics[width=0.9\linewidth]{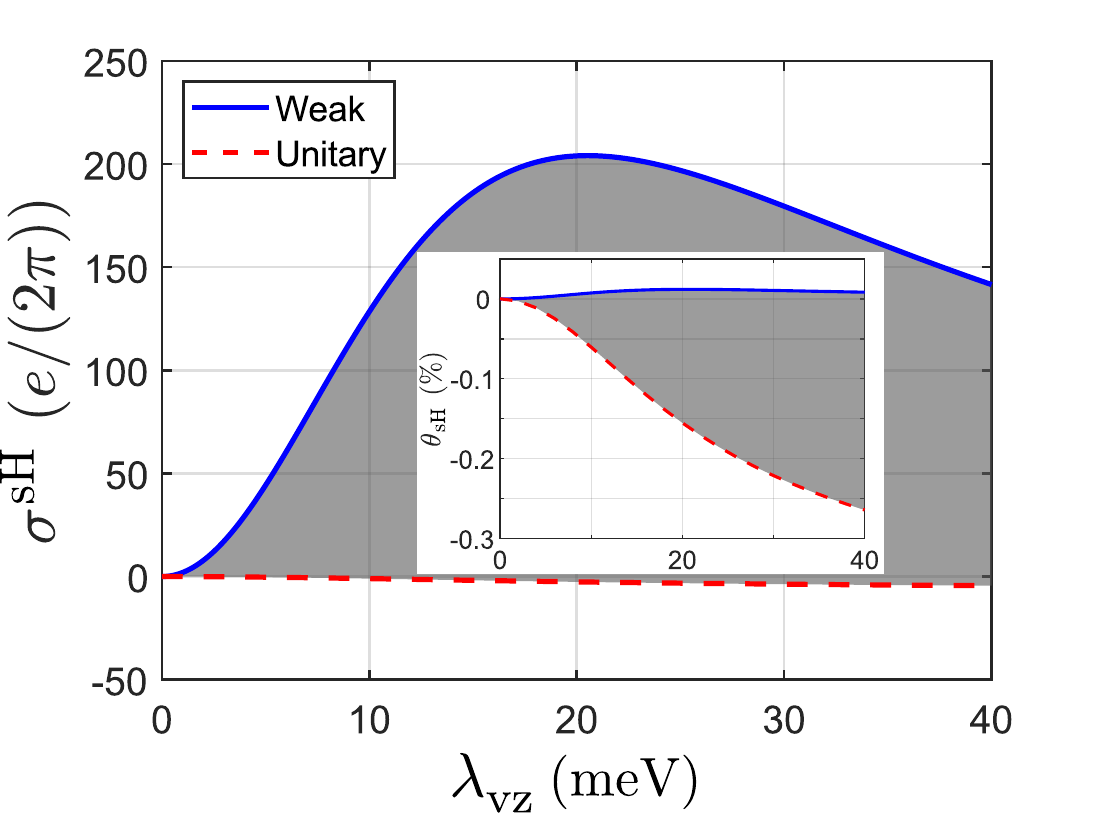}
        \caption{Valley-Zeeman coupling dependence of the spin Hall response for weak scattering potentials (Eq. \ref{SHE_Y_result}, with $u_{0} = 0.1$ eV\,nm${}^{2}$) and unitary ($u_{0} \rightarrow \infty$) limits. A fixed Fermi energy of $\varepsilon = 0.2$ eV is assumed alongside $n = 10 ^{14}$ m${}^{-2}$ and $\lambda_{\text{R}} = 20$ meV. The grey region is the area accessible with intermediate scattering strengths. Inset: Same for the spin Hall angle, $\theta_{\text{sH}}$.}
        \label{SHE_SV_plot}
    \end{figure}

    We now turn to the non-perturbative results in the scattering strength obtained by resumming the infinite  $T$-matrix series in Fig. \ref{Diagrammatics}(b) numerically. The range of impurity concentrations we focus on is chosen to yield bona fide diffusive spin   transport, i.e. $\sigma^{\text{sH}}\sim n^{-1}$.  The valley-Zeeman behavior of the steady-state spin Hall conductivity and spin Hall angle, $\theta_{\text{sH}}^{\null} = 2e\sigma^{\text{sH}}/\sigma_{xx}^{\null}$, is shown in Fig. \ref{SHE_SV_plot} in both the weak scattering and unitary limits. (For consistency, we calculate the charge conductivity, $\sigma_{xx}$, from linear response theory with the same methodology used for $\sigma^{\text{sH}}$.) Moreover, the weak-scattering limit of $\sigma^{\text{sH}}$ (solid line) is obtained via  Eq. (\ref{SHE_Y_result});  a numerical calculation in this regime is out of reach due to the smallness of the disorder self-energy. We see that while the weak scattering limit may yield a nominally large magnitude of the spin Hall response, the corresponding spin-charge conversion efficiency is significantly lower than in the unitary case ($|\theta_{\textrm{\text{sH}}}^{\textrm{unitary}}|\gg|\theta_{\textrm{\text{sH}}}^{\textrm{weak}}|$).  This can be inferred from the scaling behaviors in the perturbative regime: $\sigma^{\text{sH}}\propto u_0^{-1}$ [see Eq. (\ref{SHE_Y_result})] as opposed to the faster decay featured by the eletrical conductivity ($\sigma_{xx}\propto u_{0}^{-2}$).  Furthermore, the charge transport coefficients have distinct Fermi energy dependencies in the Born and unitary scattering regimes. Specifically, $\sigma_{xx}\sim \varepsilon^0$ (BA) and $\sigma_{xx}\sim \varepsilon^2$ (unitary) in the limit $\varepsilon \gg \Delta_s$. This is fortunate, because the measured charge conductivity in graphene-TMD closely follows the $\varepsilon^2$-law in the intermediate-to-high charge carrier density regime \cite{Benitez2020,note_unitary_limit}, thus matching the results of our theory in the unitary limit and hence evidencing its predictive power. In this strong scattering regime, not only do the predicted spin Hall angles reach detectable values (see inset to Fig. \ref{SHE_SV_plot}), more importantly, they  agree well with lateral spin Hall measurements  \cite{Benitez2020}. Additionally, $|\sigma^{\text{sH}}|$ increases with $\lambda_{\text{vz}}$ in a monotonic fashion for strong disorder, exhibiting no turning points inside a reasonable range of $\lambda_{\text{vz}}^{\null}$, unlike the weak scattering response which displays a maximum at $\lambda_{\text{vz}}^{\null} \simeq \lambda_{\text{R}}^{\null}$.
    
\begin{figure}[t]
        \centering
        \includegraphics[width=0.8\linewidth]{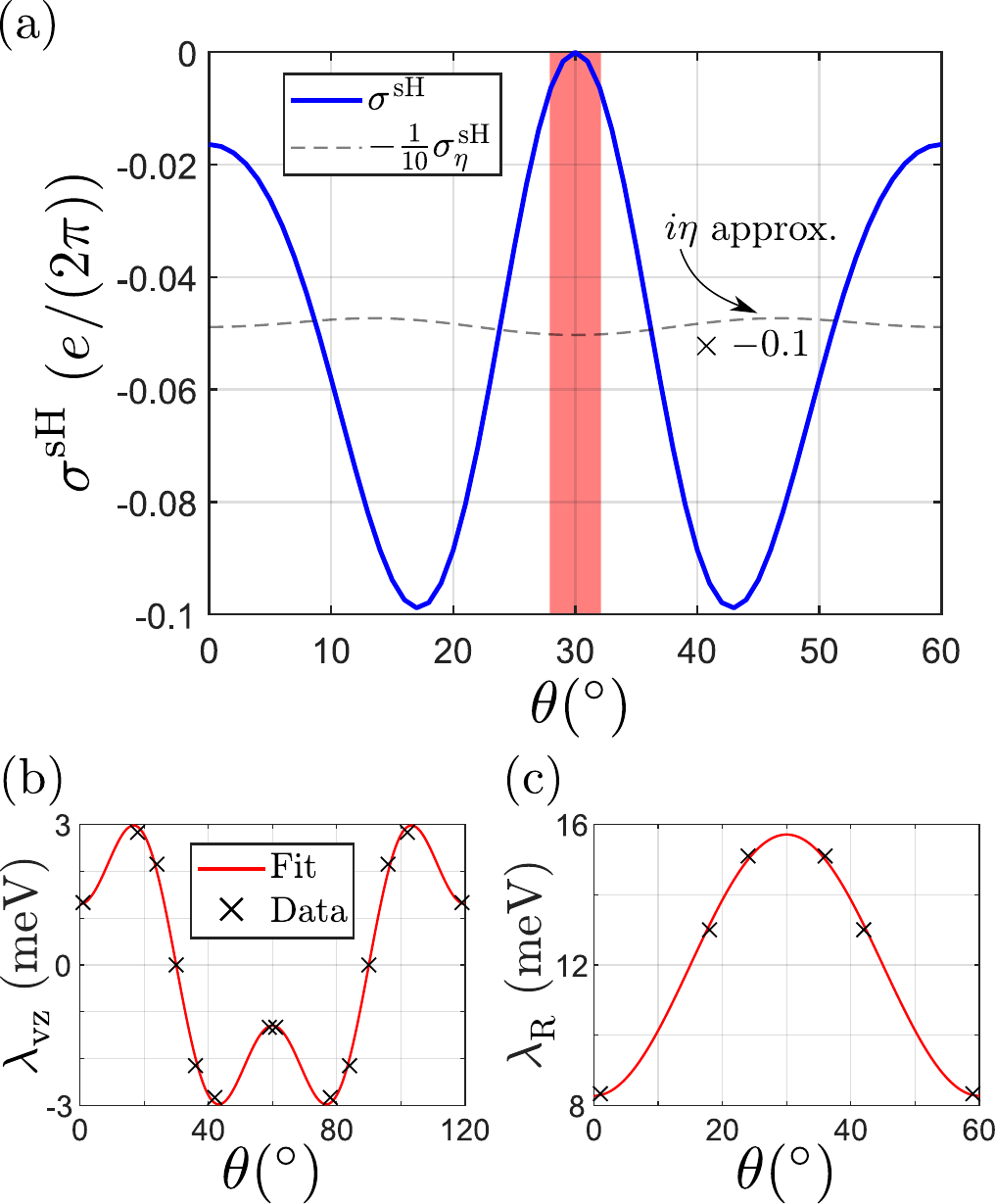}
        \caption{(a): Twist angle dependence for the renormalized and $i\eta$-approximated spin Hall conductivities for a graphene-WSe${}_{2}$ bilayer based on the experimental observations of Ref. \cite{Sun_23}, within the unitary and diffusive limits. The shaded region indicates where quantum effects will play a major role. Here we take $\varepsilon = 0.2$ eV, $n = 10^{14} \text{ m}^{-2}$, and $\Lambda = 10$ eV. (b) and (c) show minimal Fourier series fits to the experimental data (black crosses) of Ref. \cite{Sun_23}.}
          
        \label{Twist_plots}
    \end{figure}

    The considerations above show that the unitary scattering regime should be the primary focus when analyzing the SHE of realistic systems. To this end, we use the twist dependent SOC magnitudes, $\lambda_{\text{R}}(\theta)$ and $\lambda_{\text{vz}}(\theta)$, probed in recent experiments on graphene-WSe${}_{2}$ \cite{Sun_23}. To extrapolate the experimental data to twist angles greater than $\pi/6$, we exploit the twist angle symmetries of the individual SOCs \cite{David2019,Peterfalvi2022}. In practice, this is accomplished by fitting a minimal Fourier series to the   data of  Ref. \cite{Sun_23}; see Figs. \ref{Twist_plots}(b)-(c). To further improve the accuracy of our results, we also account for the SU(2)-gauge covariance breaking due to the momentum cut-off regularization ($k_{\text{max}}=\Lambda/v$,  $\Lambda$ is the energy cut-off) of our numerical scheme \cite{Supplementary_material}. The ensuing twist angle behavior of the spin Hall response in the unitary limit  is shown in Fig. \ref{Twist_plots}(a),  which is the main finding of this Letter. The significance of these results is best appreciated by a direct comparison against the $i \eta$-approximated response, $\sigma_{\eta}^{\text{sH}}$, wherein $\Sigma^{\pm} = \mp i\eta$ and vertex corrections are neglected \cite{Supplementary_material}.
   
    We immediately see that $\sigma^{\text{sH}}$ and $\sigma_{\eta}^{\text{sH}}$ differ in several ways. Most importantly,  $\sigma^{\text{sH}}$ vanishes when $\lambda_{\text{vz}}^{\null} = 0$ while $\sigma_{\eta}^{\text{sH}}$ reaches a maximal value at this point. What is more, the $i\eta$ approximation yields a response that is not only different in sign, but also an order of magnitude larger than the renormalized result. We gleam insight for this size discrepancy from the weak scattering  limit, where $\sigma^{\text{sH}} \sim \varepsilon^{-1}$ for large Fermi energies in contrast to $\sigma_{\eta}^{\text{sH}}$ tending towards some constant value \cite{Supplementary_material}. The $i \eta$ scheme irrefutably fails in modelling the SHE, even when accounting for the parametric dependencies of the broadening $\eta=\eta(\theta)$. Lastly, we note the ladder approximation, corresponding to only considering the first diagram in the skeleton expansion of Fig. \ref{Diagrammatics}(b), also fails to describe the giant skew-scattering-driven SHE modulation reported here. This is because the left-right asymmetry of scattering cross sections manifests at third-order in the scattering potential, as is well known.  Before discussing twisting effects, a cautionary remark is in order: short-range defects generate substantial valley mixing \cite{Peres_RMP_10,Ni_10,Joucken_21}, which is absent in our  disorder model. Because the skew cross section  has opposite signs in the $K$ and $K^\prime$ valleys (courtesy of the valley-Zeeman effect), intervalley scattering events diminish the skewness of spin-up/spin-down scattering channels, and thus the spin Hall conductivity \cite{Offidani_PRL_18}. Given the prominence of band-driven skew scattering in 2D materials \cite{Milletari2017}, the use of interfaces free of atomic defects  is advisable to  fully exploit the advantages of proximity-induced SOC for SHE. For a quantitative analysis of the interplay of intervalley and skew scattering  in honeycomb layered systems, we refer the reader to Refs. \cite{Milletari2017,Offidani_PRL_18} where these ideas were first discussed.
    
    %-------------------------
    %% Discussion
    %-------------------------

    \textit{Twisting effects.---}We first focus on the region of twist angles close to $30^{\circ}$ as this will be the area hosting the most exotic physics (Fig. \ref{Twist_plots}). At $\theta = 30^{\circ}$, the graphene-TMD system has $C_{3v}$ symmetry, akin to untwisted bilayers. However, unlike perfectly aligned heterostructures, there is a hidden  sublattice  symmetry in our continuum model. This is because $\lambda_{\text{vz}}(30^{\circ})=0$ and thus the clean system possesses chiral (sublattice) symmetry \cite{Schnyder_08} at zero chemical potential, $\sigma_z H_{\textbf{k}}\sigma_z=-H_{\textbf{k}}$. (We note that our model omits a small sublattice-resolved scalar potential effects in accord to perturbation theory \cite{David2019} and first principles calculations \cite{Naimer_2021}.)  As such, we speculate that, when the twist angle is equal or close to $30^\circ$, small fluctuations in the proximity-induced spin-orbit fields will dominate the SHE due to $\lambda_{\text{vz}}^{\null}$ approaching zero. These fluctuations can arise from ripples in the graphene flake \cite{Huertas-Hernando_09} and non-uniform twisting across the sample \cite{Uri_20}. Both of these will yield spatially varying spin-orbit couplings that can engender anomalous spin Hall responses \cite{Drydal2012,Huang_16}. %The macroscopic behaviour of this could be captured by treating these variations as a form of disorder in both the Rashba and valley-Zeeman couplings. 
    The exact consequence of these fluctuations makes for an interesting question for further study beyond this work. Second, we note that the spin Hall response in WSe${}_{2}$ is optimal for $\theta \approx 17^\circ$, by virtue of the maximal value of $|\lambda_{\text{vz}}(\theta)|$.  The strong modulation of $\sigma^{\text{sH}}$ demonstrated here is a direct indicator of the giant renormalization generated by the interplay of disorder and twist-dependent Fermi surface spin texture.  In closing this discussion, we finally note that changing the TMD of this heterostructure will naturally yield a different twist angle dependence of the spin Hall conductivity and angle. The resulting change cannot be predicted through intuition alone and may be drastic, as hinted at by the difference in twist angle variation of the SOC strengths shown in Ref. \cite{Peterfalvi2022}. The only guaranteed similarity is the vanishing of $\sigma^{\text{sH}}$ at $\theta = 30^{\circ}$.
  
    %-------------------------
    %% Non-local Resistances
    %-------------------------
     
     \textit{Lateral spin transport.---}Lastly, we frame our results in the context of recent experiments detecting the SHE in graphene-TMD using spin precession techniques in Hall bar geometry  \cite{Safeer2019,Benitez2020}. Within the   weak scattering regime (specifically, $u_0 \ll \nu_0(\varepsilon)^{-1}$, with $\nu_0(\varepsilon)$ the clean density of states), the observed spin Hall angles ($\theta_{\text{SHE}}^{\text{exp}} \sim$ 0.1--1 $\%$)  are not achievable, even with proximity-induced SOC choices larger than that recently mapped out by quantum interference imaging \cite{Sun_23}. Our microscopic theory   predicts  $\theta_{\text{SHE}}^{\null} \sim 0.02$\% for $\lambda_{\text{R}}^{\null} = \lambda_{\text{vz}}^{\null} = 20$ meV (see Fig. \ref{SHE_SV_plot}), indicating that the behaviour observed in spin Hall transport experiments is the result of  strong scattering potentials. Working in the unitary limit, we find that a spin Hall angle of order 0.1\% is achievable with larger SOCs or at higher impurity concentrations ($\sim 5 \times 10^{15} \text{ m}^{-2}$), however, this starts to move the system away from the diffusive limit. For example, for a system reflective of graphene-WSe${}_{2}$ \cite{Sun_23,Rao_23} ($\lambda_{\text{R}}^{\null} = 14$ meV, $\lambda_{\text{vz}}^{\null} = 3$ meV) with $n = 4.5 \times 10^{15} \text{m}^{-2}$, we obtain $\theta_{\text{SHE}}^{\null} = 0.11$\% and find $\sigma_{xx}^{\null}$ to be approximately diffusive ($\sigma_{xx}^{\null}(n)/\sigma_{xx}^{\null}(2n) = 2.3$). However, the $\sigma^{\text{sH}}$ calculated within the $T$-matrix method turns out to be non-diffusive, reflecting higher-order corrections in $n$. Given the breakdown of the diffusive limit in obtaining spin Hall angles comparable to experiment, our findings suggest that bona fide quantum effects, such as diffractive skew scattering described by crossing diagrams \cite{Milletari2016}, may play a role in the spin transport observed.

    %-------------------------
    %% Conclusions
    %-------------------------

      In conclusion, our work demonstrates the necessity for vertex corrections in the accurate modelling of the SHE in layered materials with competing broken symmetries. We find that disorder impacts pure interfacial SHEs in an unexpected way, leading to a strong oscillatory behavior of the spin Hall response upon twisting. The  twist angle dependence of the SHE uncovered here reflects the underlying quantum geometry of electronic states in regions of non-coplanar spin texture,  and allows for the regimes in which spatial fluctuations and intrinsic effects may dominate, thus raising intriguing questions for future research.
          
    %-------------------------
    %% Acknowledgements  
    %-------------------------
    %A.V., D.T.S.P., and A.F. would like to thank C.G. P\'{e}terfalvi for providing the SOC twist-dependence data.
    
    D.T.S.P., A.V., and A.F. acknowledge support from the Royal Society through Grants No. URF$\backslash$R$\backslash$191021, RF$\backslash$ERE$\backslash$210281 and RGF$\backslash$EA$\backslash$180276.
    
   \input{main.bbl}
   %\bibliography{References}

    %-------------------------
    %% Supplemental Material    
    %-------------------------
 
    \clearpage
    
    \onecolumngrid
    \section*{Supplementary Material}
    \onecolumngrid

    \section*{Neglecting vertex corrections: the $i \eta$ approximation}
    
    A common approximation in the literature is to take the self-energy to be a purely imaginary scalar quantity, $\Sigma^{\pm} = \mp i\eta$, discarding any matrix structure that may be generated by the disorder-averaging procedure. This is often used in conjunction to the neglect of vertex corrections; this combination is commonly known as the $i\eta$-approximation. Calculating the spin Hall response function within  this scheme yields
    \begin{equation}
      \sigma_{\eta}^{\text{sH}}=\frac{e}{4\pi}\frac{\lambda_{\text{R}}^{2}(\varepsilon^{2}+\lambda_{\text{vz}}^{2})}{\varepsilon^{2}(\lambda_{\text{R}}^{2}+\lambda_{\text{vz}}^{2})-\lambda_{\text{R}}^{4}-3\lambda_{\text{R}}^{2}\lambda_{\text{vz}}^{2}-\lambda_{\text{vz}}^{4}}+\mathcal{O}(\eta)\,.
        \label{SHE_no_vertex}
    \end{equation}

    The renormalized response in Eq.\,(4) of main text behaves as $\sigma^{\text{sH}} \sim \varepsilon^{-1}$ for $\varepsilon \gg \lambda_{\text{R}}, \lambda_{\text{vz}},\eta$,  whereas $\sigma_{\eta}^{\text{sH}}$ instead tends to a constant value of $e\lambda_{\text{R}}^{2}/ [4\pi (\lambda_{\text{R}}^{2}+\lambda_{\text{vz}}^{2})]$. As the Fermi energy is increased, the relative difference between the two Fermi rings of the spin-orbit coupled  bands decreases and thus would naturally yield a smaller spin Hall response, further evidencing the need for vertex corrections to capture the correct behavior of the SHE.  

    \section*{Numerical evaluation of the spin Hall conductivity}

    The use of a finite momentum cut-off in the momentum integral breaks the SU(2)-gauge covariance of the theory, and so violates Dimitrova's argument for a vanishing SHE when $\lambda_{\text{R}}\neq 0$ and $\lambda_{\text{vz}}=0$ \cite{Dimitrova2005,Milletari2017}. In calculating the spin Hall conductivity numerically for arbitrary $\lambda_{\text{R,vz}}$, we remove the (small) contribution, $\sigma^{\text{sH}}(\lambda_{\text{vz}}=0)$, due to this artificial symmetry breaking by determining the associated response for $\lambda_{\text{vz}}^{\null} = 0$, where $\sigma^{\text{sH}}_{\text{exact}}(\lambda_{\text{vz}}=0)=0$ due to the exact SU(2)-gauge covariance of the model. 
 
\end{document}

%% file: main.bbl
%apsrev4-2.bst 2019-01-14 (MD) hand-edited version of apsrev4-1.bst
%Control: key (0)
%Control: author (8) initials jnrlst
%Control: editor formatted (1) identically to author
%Control: production of article title (0) allowed
%Control: page (0) single
%Control: year (1) truncated
%Control: production of eprint (0) enabled
%